\pdfoutput=1

\documentclass[showpacs,aps,prl,twocolumn,superscriptaddress]{revtex4}
\usepackage{graphicx} 
\usepackage{dcolumn}
\usepackage{bm}
\usepackage{amssymb,amsmath}
\usepackage{epstopdf}
\usepackage[T1]{fontenc}
\usepackage[latin9]{inputenc}
\usepackage{color}
\usepackage{float}
\usepackage{graphicx}
\usepackage{esint}
\usepackage{bm}
\usepackage{graphics}
\makeatletter
\@ifundefined{textcolor}{}
{%
 \definecolor{BLACK}{gray}{0}
 \definecolor{WHITE}{gray}{1}
 \definecolor{RED}{rgb}{1,0,0}
 \definecolor{GREEN}{rgb}{0,1,0}
 \definecolor{BLUE}{rgb}{0,0,1}
 \definecolor{CYAN}{cmyk}{1,0,0,0}
 \definecolor{MAGENTA}{cmyk}{0,1,0,0}
 \definecolor{YELLOW}{cmyk}{0,0,1,0}
}

\begin{document}
\title{Collective, Coherent, and Ultrastrong Coupling of 2D Electrons with\\ Terahertz Cavity Photons}
\normalsize

\author{Qi Zhang}
\author{Minhan Lou}
\author{Xinwei Li}
\affiliation{Department of Electrical and Computer Engineering, Rice University, Houston, Texas 77005, USA}

\author{John L. Reno}
\affiliation{Sandia National Laboratories, CINT, Albuquerque, New Mexico 87185, USA}

\author{Wei Pan}
\affiliation{Sandia National Laboratories, Albuquerque, New Mexico 87185, USA}

\author{John D. Watson}
\affiliation{Department of Physics and Astronomy, Microsoft Station Q Purdue, and Birck Nanotechnology Center, Purdue University, West Lafayette, Indiana 47907, USA}

\author{Michael J. Manfra}
\affiliation{Department of Physics and Astronomy, Microsoft Station Q Purdue, and Birck Nanotechnology Center, Purdue University, West Lafayette, Indiana 47907, USA}
\affiliation{School of Materials Engineering and School of Electrical and Computer Engineering, Purdue University, West Lafayette, Indiana 47907, USA}

\author{Junichiro Kono}
\thanks{Author to whom correspondence should be addressed}
\email[]{kono@rice.edu}
\affiliation{Department of Electrical and Computer Engineering, Rice University, Houston, Texas 77005, USA}
\affiliation{Department of Physics and Astronomy, Rice University, Houston, Texas 77005, USA}
\affiliation{Department of Materials Science and NanoEngineering, Rice University, Houston, Texas 77005, USA}

\date{\today}

\begin{abstract}
Nonperturbative coupling of light with condensed matter in an optical cavity is expected to reveal a host of coherent many-body phenomena and states~\cite{CiutietAl05PRB,HeppLieb73AP,WangHioe73PRA,DeLiberato14PRL,LiberatoetAl07PRL,Moore70JMP,FullingDavies76PRSLA}.  In addition, strong coherent light-matter interaction in a solid-state environment is of great interest to emerging quantum-based technologies~\cite{ImamogluetAl99PRL,BlaisetAl04PRA}.  However, creating a system that combines a long electronic coherence time, a large dipole moment, and a high cavity quality ($Q$) factor has been a challenging goal~\cite{TodorovetAl09PRL,GeiseretAl12PRL,ScalarietAl12Science,MaissenetAl14PRB}.  Here, we report collective ultrastrong light-matter coupling in an ultrahigh-mobility two-dimensional electron gas in a high-$Q$ terahertz photonic-crystal cavity in a quantizing magnetic field, demonstrating a cooperativity of $\sim$360. The splitting of cyclotron resonance (CR) into the lower and upper polariton branches exhibited a $\sqrt{n_\mathrm{e}}$-dependence on the electron density ($n_\mathrm{e}$), a hallmark of collective vacuum Rabi splitting.  Furthermore, a small but definite blue shift was observed for the polariton frequencies due to the normally negligible $A^2$ term in the light-matter interaction Hamiltonian. Finally, the high-$Q$ cavity suppressed the superradiant decay of coherent CR, which resulted in an unprecedentedly narrow intrinsic CR linewidth of 5.6\,GHz at 2\,K.  These results open up a variety of new possibilities to combine the traditional disciplines of many-body condensed matter physics and cavity-based quantum optics.
\end{abstract}


\maketitle

Strong resonant light-matter coupling in a cavity setting is an essential ingredient in fundamental cavity quantum electrodynamics (QED) studies~\cite{Haroche13RMP} as well as in cavity-QED-based quantum information processing~\cite{ImamogluetAl99PRL,BlaisetAl04PRA}. In particular, a variety of {\em solid-state} cavity QED systems have recently been examined~\cite{KhitrovaetAl06NP,TormaBarnes15RPP,TabuchietAl15Science,ViennotetAl15Science}, not only for the purpose of developing scalable quantum technologies, but also for exploring novel many-body effects inherent to condensed matter. For example, collective $\sqrt{N}$-fold enhancement of light-matter coupling in an $N$-body system~\cite{Dicke54PR}, combined with colossal dipole moments available in solids, compared to traditional atomic systems, is promising for entering uncharted regimes of ultrastrong light-matter coupling.  Nonintuitive quantum phenomena can occur in such regimes, including a ``squeezed'' vacuum state~\cite{CiutietAl05PRB}, the Dicke superradiant phase transition~\cite{HeppLieb73AP,WangHioe73PRA}, the breakdown of the Purcell effect~\cite{DeLiberato14PRL}, and quantum vacuum radiation~\cite{LiberatoetAl07PRL} induced by the dynamic Casimir effect~\cite{Moore70JMP,FullingDavies76PRSLA}.

Specifically, in a cavity QED system, there are three rates that jointly characterize different light-matter coupling regimes: $g$, $\kappa$, and $\gamma$. The parameter $g$ is the coupling constant, with 2$g$ being the vacuum Rabi splitting between the two normal modes, the lower polariton (LP) and upper polariton (UP), of the coupled system. The parameter $\kappa$ is the photon decay rate of the cavity; $\tau_\mathrm{cav}$ $=$ $\kappa^{-1}$ is the photon lifetime of the cavity, and the cavity $Q$ $=$ $\omega_0 \tau_\mathrm{cav}$ at mode frequency $\omega_0$.  The parameter $\gamma$ is the nonresonant matter decay rate, which is usually the decoherence rate in the case of solids.  Strong coupling is achieved when the splitting, $2g$, is much larger than the linewidth, ($\kappa + \gamma)/2$, and  ultrastrong coupling is achieved when $g$ becomes a considerable fraction of $\omega_{0}$. The two standard figures of merit to measure the coupling strength are $C$ $\equiv$ $4g^2/(\kappa\gamma)$ and $g/\omega_0$; here, $C$ is called the {\em cooperativity} parameter~\cite{ViennotetAl15Science}, which is also the determining factor for the onset of optical bistability through resonant absorption saturation~\cite{BonifacioLugiato82Book}.  In order to maximize $C$ and $g/\omega_0$, one should construct a cavity QED setup that combines a large dipole moment (i.e., large $g$), a small decoherence rate (i.e., small $\gamma$), a large cavity $Q$ factor (i.e., small $\kappa$), and a small resonance frequency $\omega_0$.

III-V semiconductor quantum wells (QWs) provide one of the cleanest and most tunable solid-state environments with quantum-designable optical properties. Microcavity QW-exciton-polaritons represent a landmark realization of a strongly coupled light-condensed-matter system that exhibits a rich variety of coherent many-body phenomena~\cite{DengetAl10RMP}.  However, the large values of $\omega_0$ and relatively small dipole moments for interband transitions make it impractical to achieve large values of $g/\omega_0$ using exciton-polaritons.  Intraband transitions, such as intersubband transitions (ISBTs)~\cite{CiutietAl05PRB} and cyclotron resonance (CR)~\cite{HagenmulleretAl10PRB}, are much better candidates for accomplishing ultrastrong coupling because of their small $\omega_0$, typically in the midinfared and terahertz (THz) range, and their enormous dipole moments (10s of $e$-\AA). Experimentally, ultrastrong coupling has indeed been achieved in GaAs QWs using ISBTs~\cite{TodorovetAl09PRL,GeiseretAl12PRL} and CR~\cite{ScalarietAl12Science,MaissenetAl14PRB}. In the latter case, a record high value of $g/\omega_0$ $=$ 0.87 has been reported~\cite{MaissenetAl14PRB}.  In all these previous intraband studies of ultrastrong light-matter coupling, however, due to ultrafast decoherence (large $\gamma$) and/or lossy cavities (large $\kappa$), the value of $C$ remained small, i.e., the standard strong-coupling criterion ($C \gg 1$) was not satisfied.

Here, we simultaneously achieved small $\gamma$ and small $\kappa$ in ultrahigh-mobility two-dimensional electron gases (2DEGs) in GaAs QWs placed in a high-$Q$ 1D THz photonic-crystal cavity (PCC) in a perpendicular magnetic field. We achieved ultrastrong coupling ($C > 300$ and $g/\omega_0 \sim 0.1$) between coherent CR and THz cavity photons, observing vacuum Rabi splitting (Rabi oscillations) in the frequency (time) domain. 
Furthermore, we observed a $\sqrt{n_\mathrm{e}}$-dependence of 2$g$ on the electron density ($n_\mathrm{e}$), signifying the collective nature of light-matter coupling~\cite{Dicke54PR,KaluznyetAl83PRL,AmsussetAl11PRL,TabuchietAl14PRL,ZhangetAl14PRL2}. A value of $g/\omega_0$ $=$ 0.12 was obtained with just a single QW with a moderate $n_\mathrm{e}$. Finally, the previously identified superradiant decay of CR in high-mobility 2DEGs~\cite{ZhangetAl14PRL} was significantly suppressed by the presence of the high-$Q$ THz cavity. As a result, we observed ultranarrow polariton lines, yielding an intrinsic CR linewidth as small as 5.6\,GHz (or a CR decay time of 57\,ps) at 2\,K. 

\begin{figure}
\includegraphics[scale = 0.55]{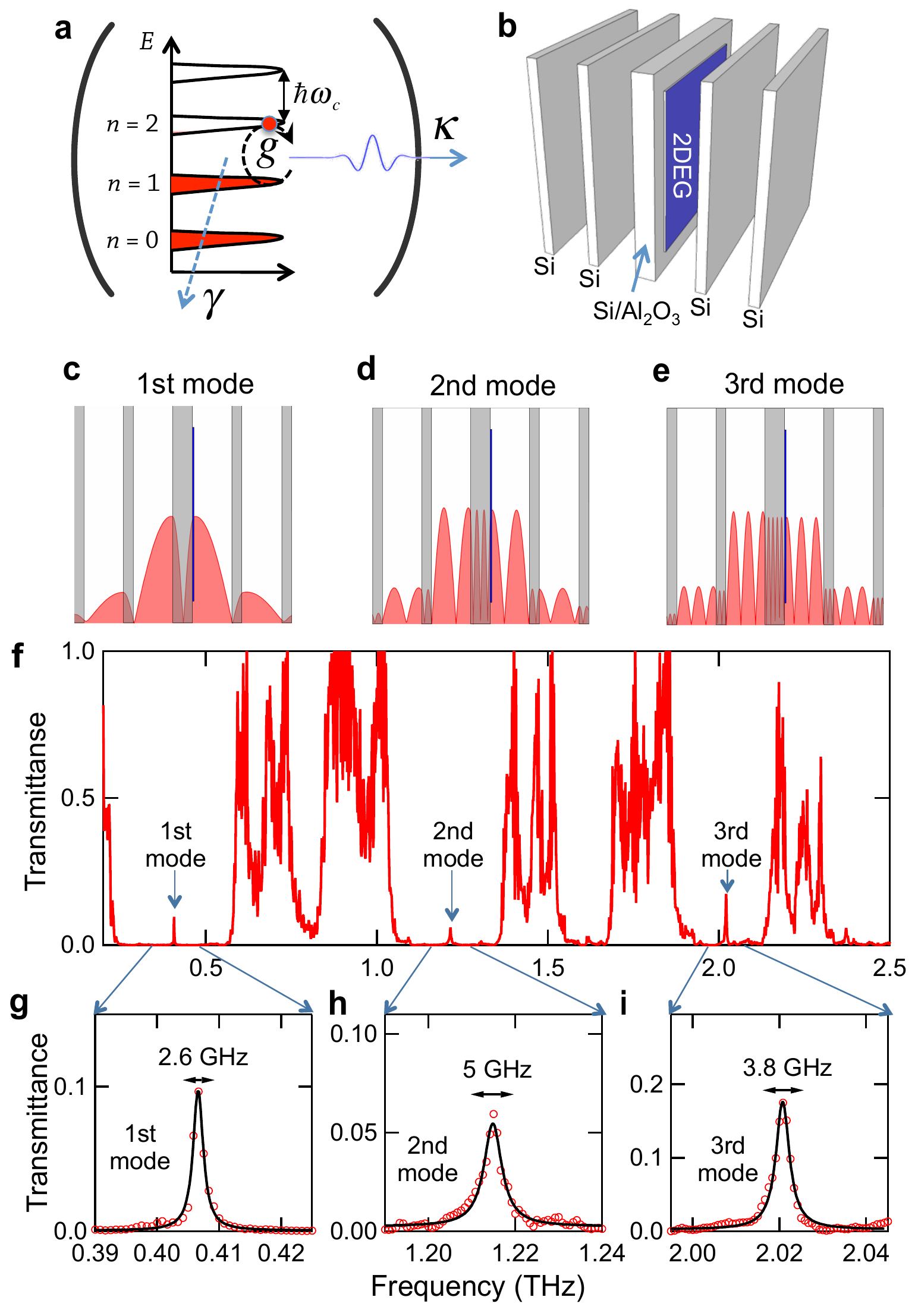}
\caption{\textbf{1D THz photonic crystal cavity (PCC) with a high-mobility 2DEG.} \\
\textbf{a},~Schematic diagram for cyclotron resonance involving two adjacent Landau levels resonantly coupled with a THz cavity field. $g$: light-matter coupling constant, $\kappa$: photon decay rate, and $\gamma$: matter decay rate. \, \textbf{b},~1D THz PCC structure. Two silicon layers are placed on each side of the center defect layer. The blue part is the transferred 2DEG thin film. \, \textbf{c}, \textbf{d}, and \textbf{e},~Calculated electric field amplitude distribution inside the cavity for the 1st, 2nd, and 3rd cavity modes. The 2DEG is located at the field maximum for all three cavity modes. \, \textbf{f},~Experimental power transmittance spectrum for the cavity.  Three sharp cavity modes are clearly resolved in the middle of each stop-band. \, \textbf{g}, \textbf{h}, and \textbf{i},~Zoom-in spectra for the three cavity modes, together with Lorentzian fits. The FWHM ($Q$) values are 2.6\,GHz (150), 5\,GHz (243), and 3.8\,GHz (532) for the 1st, 2nd, and 3rd modes, respectively.}
\end{figure}

High-mobility GaAs 2DEG samples were studied using THz time-domain magnetospectroscopy. The magnetic field quantized the density of states of the 2DEG into Landau levels. As schematically shown in Fig.\,1\textbf{a}, THz cavity photons are coupled with the transition between adjacent Landau levels, i.e., CR. Figure~1\textbf{b} shows our 1D THz PCC design, consisting of two layers of 50-$\mu$m-thick undoped Si wafers on each side as a Bragg mirror.  Thanks to the large contrast of refractive index between Si (3.42 in the THz range) and vacuum, only a few layers of Si were required to achieve sufficient cavity confinement of THz radiation with high $Q$ values.  
A substrate-removed 4.5-$\mu$m-thick GaAs 2DEG sample was placed on the central ``defect'' layer of the PCC, which was a 100-$\mu$m-thick Si (sapphire) wafer in Cavity~1 (Cavity~2).  
Calculated electric field distribution inside Cavity~1 is shown in Fig.\,1, \textbf{c}, \textbf{d}, and \textbf{e}, for the first, second, and third cavity modes, respectively.  The spatial overlap of the 2DEG and the electric field maximum ensured the strongest light-matter coupling. 

Figure~1\textbf{f} shows a THz transmission spectrum for Cavity 1, containing a 2DEG, at 4\,K.  Three photonic band gaps are seen as transmission stop-bands.  At the center of each stop-band, a sharp cavity mode is observed. As shown in Fig.\,1, \textbf{g}, \textbf{h}, and \textbf{i}, the full-width-at-half-maximum (FWHM) values, or $\kappa$, of these cavity modes were 2.7\,GHz, 5.0\,GHz, and 3.8\,GHz, 
corresponding to $Q$ factors of 150, 243, and 532, respectively; note that these numbers are slightly lower than those for an empty cavity without including the 2DEG, which were 183, 450, and 810.  These $Q$-factors are one to two orders of magnitude higher than those reported for the THz metamaterial resonators employed in previous untrastrong-coupling studies using 2DEG CR~\cite{ScalarietAl12Science,MaissenetAl14PRB}.  
In the following, experimental data recorded with Cavity~1 are shown.  

By varying the magnetic field ($B$), we continuously changed the detuning between the cyclotron frequency ($\omega_\mathrm{c}$ $=$ $eB/m^*$, where $m^*$ $=$ 0.07$m_\mathrm{e}$ is the electron effective mass of GaAs  and $m_\mathrm{e}$ $=$ 9.11 $\times$ 10$^{-11}$\,kg) and the cavity mode frequency ($\omega_0$): $\Delta \equiv \omega_0 - \omega_\mathrm{c}$.
Clear anticrossing behavior, expected for strong coupling, is shown in Fig.\,2\textbf{a} for the first cavity mode in Cavity~1. Two polariton branches (LP and UP) were formed through the hybridization of CR and the THz cavity photons.  The central peak originates from the transmission of the CR-inactive circular-polarization component of the linearly polarized incident THz beam, which does not interact with the 2DEG and whose position is independent of $B$. The FWHM of the central peak is thus given by $\kappa$, while that for the LP and UP peaks at $\Delta$ $=$ 0 is given by $(\kappa + \gamma)/2$.  Therefore, from the $\Delta$ $=$ 0 spectrum ($B_\mathrm{r}$ $=$ 1.00\,T for this mode), we determined 
$(2g,\kappa,\gamma,\omega_0)/2\pi$ $=$ (74, 2.6, 5.6, 407)GHz, yielding $C$ $=$ 360 and $g/\omega_0$ $=$ 0.09.  
Parameter values determined in this manner for all modes in both Cavities 1 and 2 are summarized in Table~\ref{Parameters}, together with cavity parameters and resonance conditions.

\begin{table*}
  \centering
  \begin{tabular}{|c|c|c|c|c|c|c|c|c|c|c|}
    \hline 
    Cavity & Mode & $Q$ & $\tau_\mathrm{cav}$ & $B_\mathrm{r}$ & $\nu$ & $\omega_0/2\pi$ & $2g/2\pi$ & $\kappa/2\pi$ & $\gamma/2\pi$ & $g/\omega_0$\\
     &  &  & (ps) & (T) &  & (GHz) & (GHz) & (GHz) & (GHz) &\\
    \hline
    \hline
	1 & 1  & 183 & 69 & 1.000 & 12.4  & 407 & 74 & 2.6 & 5.6 & 0.09\\
	\hline
	1 & 2  & 450 & 57 & 2.975 & 4.2 & 1218 & 66 & 5.0 & NA$^\dagger$ & 0.03 \\
	\hline
	1 & 3 & 810 & 61 & 4.925 & 2.5 & 2020 & 60 & 3.8 & NA$^\dagger$ & 0.015\\
	\hline
	2 & 1  &  $>$40$^*$& -- & 0.950 & 13  & 375 & 90 & $<$10$^*$ & $<$10$^*$ & 0.12\\
	\hline
	2 & 2  & $>$110$^*$ & -- & 2.700 & 4.6 & 1100 & 90 & $<$10$^*$ & $<$10$^*$ & 0.04\\
	\hline
  \end{tabular}
  \caption{Cavity and matter parameters extracted from experimental data for each cavity mode in the two cavities used.  $Q$:\,quality factor of cavity without 2DEG, $\tau_\mathrm{cav}$:\,photon lifetime of cavity without 2DEG, $B_\mathrm{r}$:\,resonance magnetic field, $\nu$ $=$ $n_\mathrm{e}h/eB$:\,Landau-level filling factor at $B$ $=$ $B_\mathrm{r}$, $\omega_0$:\,mode frequency, 2$g$:\,vacuum Rabi splitting, $\kappa$:\,photon decay rate (with 2DEG), and $\gamma$:\,nonradiative matter decay rate. $^*$Instrument limited due to insufficient time-scan range. $^\dagger$Not accurately obtainable because of distorted lineshape.}
  \label{Parameters}
\end{table*}

\begin{figure}
\includegraphics[scale = 0.57]{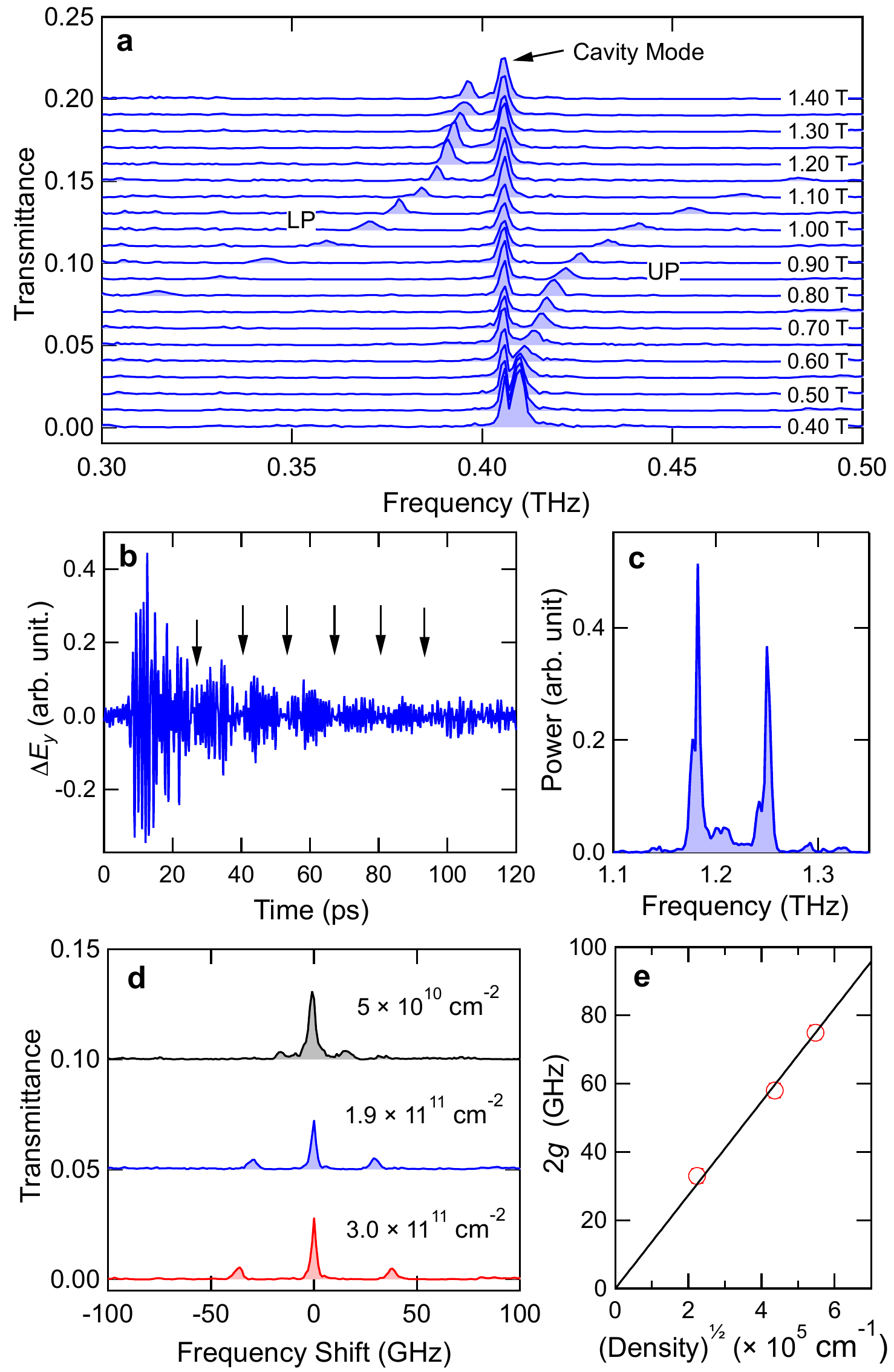}
\caption{\textbf{Observation of collective ultrastrong light-matter coupling in a 2D electron gas in a THz photonic-crystal cavity.} \,
\textbf{a},~Anticrossing of cyclotron resonance (CR) and the first cavity mode, exhibiting the lower-polariton (LP) and upper-polariton (UP) branches. The central peak due to the cavity mode results from the CR inactive circularly polarized component of the linearly polarized THz beam. Transmission spectra at different magnetic fields are vertically offset for clarity. The magnetic field increases from 0.4\,T (bottom) to 1.4\,T (top). \, \textbf{b},~Vacuum Rabi oscillations in the time domain. CR is resonantly coupled with the second cavity mode at 2.975\,T. $\Delta E_{y}$ $=$ $E_y(+\textrm{2.975\,T})-E_y(-\textrm{2.975\,T})$ is the measured difference between the transmitted THz waveforms taken at +2.975\,T and $-$2.975\,T in the $y$-polarization direction. The residual CR inactive cavity mode was removed by a numerical notch filter. The beating nodes of the two polaritons are indicated by arrows. \, \textbf{c},~The frequency-domain spectrum of $\Delta E_{y}$ in \textbf{b}.  \, \textbf{d},~Vacuum Rabi splitting observed for 2DEGs with three different electron densities.  CR was resonant with the fundamental cavity mode. \, \textbf{e},~Square root of $n_\textrm{e}$ dependence of vacuum Rabi splitting, evidencing the collective nature of light-matter coupling.}
\end{figure}

As in other cavity QED systems based on atoms and microcavity excitons, vacuum Rabi splitting in the frequency domain can be directly observed as time-domain oscillations~\cite{KaluznyetAl83PRL,NorrisetAl94PRB,ZhangetAl14PRL2}.  Experimentally, for an incident THz beam linearly polarized in the $x$ direction, we measured the $y$-polarization component, $E_{y}$, of the transmitted THz wave, in both positive ($+B$) and negative ($-B$) fields, and took the difference $\Delta E_{y}$ $=$ $E_y(+B)-E_y(-B)$, to eliminate any background noise. The CR inactive mode was numerically filtered out.  As shown in Fig.\,2\textbf{b}, 
the measured $\Delta$$E_{y}$ signal showed strong beating between the two polariton modes, which can be viewed as coherent repetitive energy exchange between the matter resonance and the cavity photons. At each beating node (indicated by an arrow), energy is stored in the 2DEG CR. The average time separation between two adjacent beating nodes was about 13-15\,ps, matching the $2g$ splitting in the frequency domain; see also Fig.\,2\textbf{c} for the Fourier transform of the time-domain oscillations. The beating lasts for dozens of picoseconds, indicating a long intrinsic CR coherence time.

In analogy to the physics of many-atom light-matter interactions~\cite{Dicke54PR}, one crucial question is whether the Rabi splitting observed here is a fully coherent behavior of a large number of individual electrons in the 2DEG. Figure~2, \textbf{d} and \textbf{e}, show three spectra showing polariton manifestation at $\Delta$ $=$ 0 for different electron densities ($n_\mathrm{e}$), when CR is in resonance with the first cavity mode in Cavity~1. The vacuum Rabi splitting ($2g$) between the LP and UP peaks exhibited a square-root dependence on $n_\mathrm{e}$ (Fig.\,2\textbf{e}), which is strong evidence for {\em collective} vacuum Rabi splitting, as observed in atomic gases~\cite{KaluznyetAl83PRL} and spin ensembles~\cite{AmsussetAl11PRL,TabuchietAl14PRL,ZhangetAl14PRL2}. This observation validates the notion that billions of 2D electrons are interacting with a common cavity THz photon field in a fully coherent manner. By extrapolation, the vacuum Rabi splitting for CR of a single electron is estimated to be 0.14\,MHz. It is also worth noting that we should be able to increase the vacuum Rabi splitting further by using multiple layers of a 2DEG with a higher electron density.

The coupled system of Landau-quantized 2D electrons and THz cavity photons can be described by the following Hamiltonian~\cite{HagenmulleretAl10PRB}: $\hat{H}_\mathrm{tot}$ $=$ $\hat{H}_\mathrm{CR} + \hat{H}_\mathrm{cavity} + \hat{H}_\mathrm{int} + \hat{H}_\mathrm{dia}$, where $\hat{H}_\mathrm{CR}$ $=$ $\hbar\omega_\mathrm{c}\hat{b}^\dag\hat{b}$, $\hat{H}_\mathrm{cavity}$ $=$ $\hbar\omega_{0}\hat{a}^\dag\hat{a}$, $\hat{H}_\mathrm{int}$ $=$ $\hbar{g}\hat{a}(\hat{b}^\dag - \hat{b}) + \hbar{g} \hat{a}^\dag (\hat{b}^\dag - \hat{b})$, and $\hat{H}_\mathrm{dia}$ $=$ $(\hbar{g}^2/\omega_\mathrm{c})(\hat{a}^\dag \hat{a}^\dag + \hat{a}^\dag \hat{a} + \hat{a}\hat{a}^\dag + \hat{a}\hat{a})$.  The first two terms, $\hat{H}_\mathrm{cavity}$ and $\hat{H}_\mathrm{CR}$, represent, respectively, the energy of the cavity mode at $\omega_0$ and the energy of the 2DEG in a $B$ with frequency $\omega_\mathrm{c}$. The operators $\hat{a}$ and $\hat{a}^\dag$ ($\hat{b}$ and $\hat{b}^\dag$) are the annihilation and creation operators for cavity photons (collective CR excitations), respectively. The light-matter interaction term, $\hat{H}_\mathrm{int}$, with coupling strength $g$ includes counter-rotating terms, $\hbar{g}(\hat{a}^\dag\hat{b}^\dag - \hat{a}\hat{b})$, which are usually neglected under the rotating-wave approximation.  Also included in the total Hamiltonian is the diamagnetic term, $\hat{H}_\mathrm{dia}$, also known as the $A^2$ term because, mathematically, it is proportional to the square of the vector potential $A$ of the light field.  The pre-factor $\hbar g^2/\omega_\mathrm{c}$ of the $A^2$ term suggests that this term is negligible in the weak-coupling regime but can have measurable effects in the ultrastrong-coupling regime where $g$ is on the order of $\omega_\mathrm{c}$.

\begin{figure}
\includegraphics[scale = 0.7]{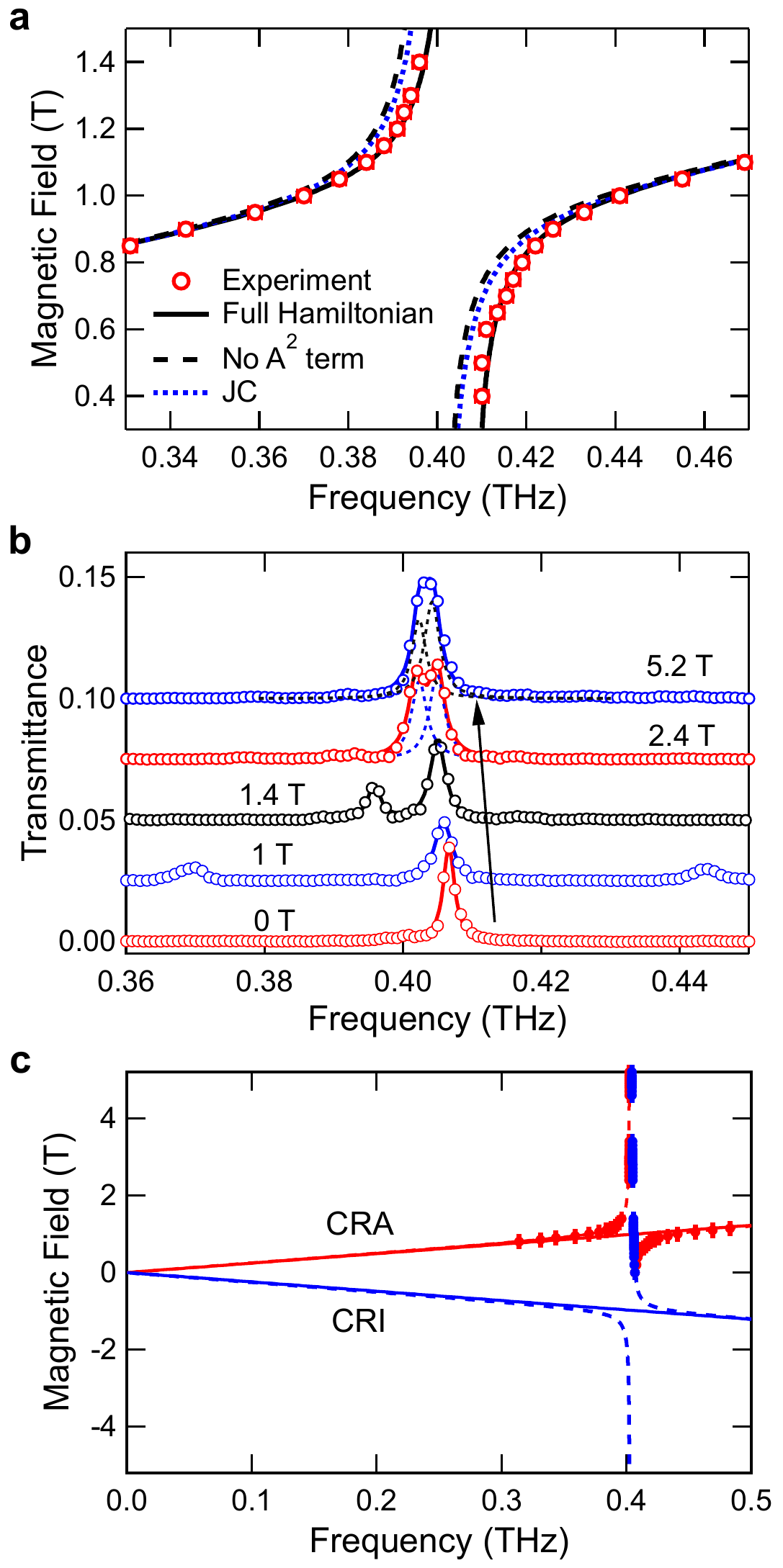}
\caption{\textbf{Theoretical modeling of ultrastrong THz-light-matter coupling.} \,
\textbf{a},~The best fits to our experimental data for the first mode using three different Hamiltonians: i)~the full CR-cavity Hamiltonian (black solid), ii)~the full CR Hamiltonian without the $A^2$ term (black dashed), and iii)~the Jaynes-Cummings Hamiltonian (blue dotted). Only the full CR-cavity Hamiltonian reproduced the experimental data well, suggesting the non-negligible contribution of the $A^2$ term on the polariton frequencies, i.e., the system is in the ulrastrong coupling regime. The best fit was achieved when $g/\omega_{0}$ $=$ 0.09 for Cavity~1. \, \textbf{b},~Power transmittance spectra at specific magnetic fields. Lorentzian fits are shown as solid lines. The residual cavity mode exhibits a red shift with increasing magnetic field, which is due to the coupling with the CR inactive mode at a negative frequency. \, \textbf{c},~Peak positions of the polariton modes (red solid diamonds) and the residual cavity mode (blue solid diamonds) as a function of magnetic field. The blue and red solid lines are CR active (CRA) and inactive (CRI) modes, respectively. Quantum mechanical fitting curves for polariton branches are shown as dashed lines.}
\end{figure}

Figure~3\textbf{a} presents the best fits to our data with three different Hamiltonians: i)~$\hat{H}_\mathrm{CR} + \hat{H}_\mathrm{cavity} + \hat{H}_\mathrm{int} + \hat{H}_\mathrm{dia}$ (full Hamiltonian), ii)~$\hat{H}_\mathrm{CR} + \hat{H}_\mathrm{cavity} + \hat{H}_\mathrm{int}$, and iii)~$\hat{H}_\mathrm{CR} + \hat{H}_\mathrm{cavity} + \hbar{g}(\hat{a}\hat{b}^\dag + \hat{a}^\dag \hat{b})$ (the Jaynes-Cummings Hamiltonian, $\hat{H}_\mathrm{JC}$). At the optimum fitting, the value of $g/\omega_{0}$ was determined to be 0.09 (for Cavity~1). The full CR-cavity Hamiltonian, i), provided the best fit, while the other two failed to show the non-negligible blue shift of the polariton modes. With $g/\omega_{0}$ $=$ 0.1, the contribution of the $A^2$ term is expected to be on the order of 0.01$\omega_0$. This 1\% contribution from the $A^2$ term indeed explains the observed blue shift, which is another manifestation of the fact that our system is in the ultrastrong-coupling regime.
Furthermore, even the cavity mode is affected by ultrastrong coupling: as shown in Fig.\,3\textbf{b}, the residual cavity mode exhibits a red shift with increasing magnetic field. This unexpected behavior occurs because the ultrastrong light-matter interaction even enables coupling between the cavity mode and the CR inactive mode at a negative frequency. The cavity mode is actually part of the UP branch for the CR inactive mode, as shown in Fig.\,3\textbf{c}.

Finally, we studied the extracted value of $\gamma$ as a function of temperature.  It has previously been shown that the decay of CR in free space is dominated by collective radiative decay, or superradiance, in ultrahigh-mobility 2DEGs, showing a decay rate that is proportional to $n_\mathrm{e}$~\cite{ZhangetAl14PRL}.  This radiative decay mechanism is very strong and dominant at low temperatures, faster than any other phase breaking scattering processes; thus, CR lines are much broader than expected from the sample mobility.  In the present case, however, the emitted coherent CR radiation cannot readily escape from the high-$Q$ cavity and thus re-excites coherent CR {\em multiple times}.  Hence, this reversible emission and absorption in a strongly coupled cavity-2DEG system strongly suppresses the superradiant decay, revealing the intrinsic CR decoherence rate, $\Gamma_\mathrm{CR}$ (s$^{-1}$) $=$ $(\pi\tau_\mathrm{CR})^{-1}$. This dramatic suppression of radiative decay is opposite to the Purcell effect in a high-$Q$ cavity expected in the weak-coupling regime and can only be understood within the framework of strong coupling~\cite{TignonetAl95PRL}, where superradiant decay is suppressed by the reversible absorption and emission processes.

\begin{figure*}
\includegraphics[scale = 0.56]{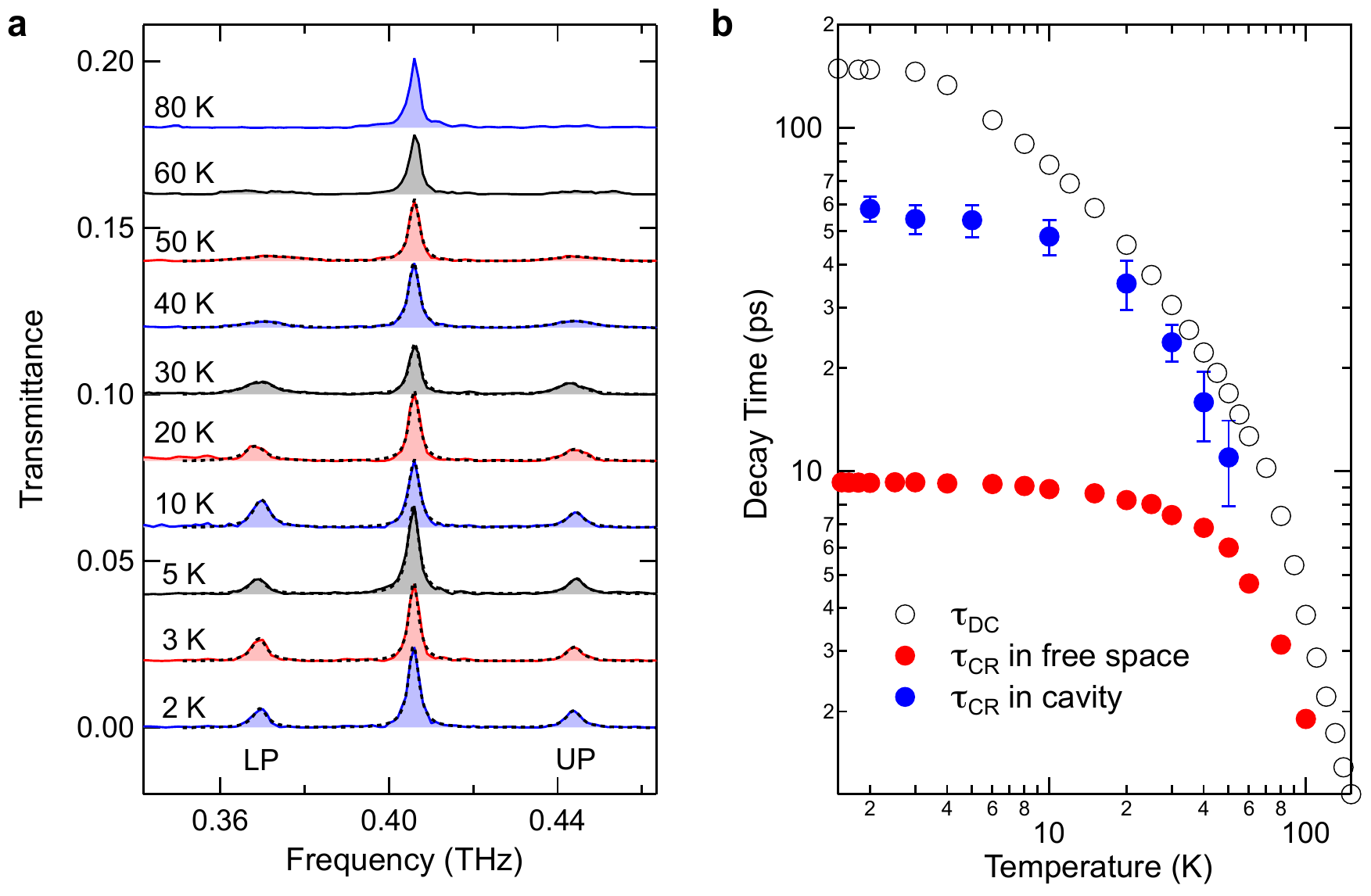}
\caption{\textbf{Observation of ultranarrow CR linewidths due to the suppression of superradiant decay by the high-$Q$ cavity.} \,
\textbf{a},~Temperature dependent spectra showing polariton peaks at zero detuning with the fundamental cavity mode from 2\,K to 80\,K with Lorentzian fits (black dashed lines). Traces are vertically offset for clarity. \, \textbf{b},~Temperature dependence of the CR decay time $\tau_\mathrm{CR}$ measured in free space (red solid circles), $\tau_\mathrm{CR}$ measured in the cavity (blue solid circles), and the DC momentum scattering time $\tau_\mathrm{DC}$ (black open circles). The value of $\tau_\mathrm{CR}$ in the cavity is 57\,ps at 2\,K, significantly enhanced from the superradiance-limited $\tau_\mathrm{CR}$ of 10\,ps in free space.}
\end{figure*}

Figure~4\textbf{a} presents temperature-dependent transmission spectra at $\Delta = 0$ for the first cavity mode in Cavity~1. The LP and UP peaks significantly broadens above 20\,K, becoming unobservable above 80\,K. The central peak, on the other hand, remains essentially unchanged as the temperature increases, serving as an excellent linewidth reference. At each temperature, we determined both $\kappa$ and $\gamma$, using the procedure described earlier (Fig.\,2\textbf{a}).  
As shown in Fig.\,4\textbf{b}, the CR decay time in the cavity is 57~$\pm~4$\,ps at 2\,K, much longer than the superradiance-limited value in free space (10\,ps). Therefore, the CR decay time measured in our high-$Q$ cavity is the intrinsic CR decay time, $\tau_\mathrm{CR}$, due to nonradiative decay mechanisms (i.e., scattering). Above 20\,K, $\tau_\mathrm{CR}$ approaches the DC momentum scattering time, $\tau_\mathrm{DC}$, where piezoelectric scattering and polar optical phonon scattering dominate. At 2\,K, $\tau_\mathrm{CR}$ is still lower than $\tau_\mathrm{DC}$.  How the CR linewidth in a high-mobility 2DEG changes with the magnetic field and temperature is a long-standing question~\cite{AndreevetAl14APL}, 
and a systematic study of `superradiance-free' CR widths should provide significant new insight.  

We have demonstrated collective and ultrastrong light-matter coupling between the CR of a 2DEG and THz cavity photons in a high-$Q$ 1D photonic-crystal cavity, with a cooperativity up to 360. 
Unlike the near-field coupling of metamaterial resonators, our THz cavity scheme is applicable to both 2D and bulk materials, which will allow us to study various strongly correlated systems with collective many-body excitations in the THz range, e.g., magnetically ordered systems, high-temperature superconductors, and heavy-fermion systems. 
Hence, our demonstration of ultrastrong light-matter interaction in high-$Q$ THz cavities opens a door to a plethora of new possibilities to combine the traditional disciplines of many-body condensed matter physics and quantum optics of cavity QED.

\section*{Acknowledgments}

We thank Andrey Chabanov for useful discussions on cavity design.  J.K.\ acknowledges support from the National Science Foundation (Grant No.\ DMR-1310138). This work was performed, in part, at the Center for Integrated Nanotechnologies, a U.S.\ Department of Energy, Office of Basic Energy Sciences user facility. Sandia National Laboratories is a multiprogram laboratory managed and operated by Sandia Corporation, a wholly owned subsidiary of Lockheed Martin Corporation, for the U.S.\ Department of Energy's National Nuclear Security Administration under Contract No.\ DE-AC04-94AL85000. The work at Sandia was supported by the U.S.\ Department of Energy, Office of Science, Materials Sciences and Engineering Division. Growth and characterization completed at Purdue by J.D.W.\ was supported by the Department of Energy, Office of Basic Energy Sciences, Division of Materials Sciences and Engineering under Award No.\ DE-SC0006671. M.J.M.\ acknowledges additional support from the W.\ M.\ Keck Foundation and Microsoft Research.





\begin{thebibliography}{30}
\expandafter\ifx\csname natexlab\endcsname\relax\def\natexlab#1{#1}\fi
\expandafter\ifx\csname bibnamefont\endcsname\relax
  \def\bibnamefont#1{#1}\fi
\expandafter\ifx\csname bibfnamefont\endcsname\relax
  \def\bibfnamefont#1{#1}\fi
\expandafter\ifx\csname citenamefont\endcsname\relax
  \def\citenamefont#1{#1}\fi
\expandafter\ifx\csname url\endcsname\relax
  \def\url#1{\texttt{#1}}\fi
\expandafter\ifx\csname urlprefix\endcsname\relax\def\urlprefix{URL }\fi
\providecommand{\bibinfo}[2]{#2}
\providecommand{\eprint}[2][]{\url{#2}}

\bibitem[{\citenamefont{Ciuti et~al.}(2005)\citenamefont{Ciuti, Bastard, and
  Carusotto}}]{CiutietAl05PRB}
\bibinfo{author}{\bibfnamefont{C.}~\bibnamefont{Ciuti}},
  \bibinfo{author}{\bibfnamefont{G.}~\bibnamefont{Bastard}}, \bibnamefont{and}
  \bibinfo{author}{\bibfnamefont{I.}~\bibnamefont{Carusotto}},
  \bibinfo{journal}{Phys. Rev. B} \textbf{\bibinfo{volume}{72}},
  \bibinfo{pages}{115303} (\bibinfo{year}{2005}).

\bibitem[{\citenamefont{Hepp and Lieb}(1973)}]{HeppLieb73AP}
\bibinfo{author}{\bibfnamefont{K.}~\bibnamefont{Hepp}} \bibnamefont{and}
  \bibinfo{author}{\bibfnamefont{E.~H.} \bibnamefont{Lieb}},
  \bibinfo{journal}{Ann. Phys.} \textbf{\bibinfo{volume}{76}},
  \bibinfo{pages}{360} (\bibinfo{year}{1973}).

\bibitem[{\citenamefont{Wang and Hioe}(1973)}]{WangHioe73PRA}
\bibinfo{author}{\bibfnamefont{Y.~K.} \bibnamefont{Wang}} \bibnamefont{and}
  \bibinfo{author}{\bibfnamefont{F.~T.} \bibnamefont{Hioe}},
  \bibinfo{journal}{Phys. Rev. A} \textbf{\bibinfo{volume}{7}},
  \bibinfo{pages}{831} (\bibinfo{year}{1973}).

\bibitem[{\citenamefont{De~Liberato}(2014)}]{DeLiberato14PRL}
\bibinfo{author}{\bibfnamefont{S.}~\bibnamefont{De~Liberato}},
  \bibinfo{journal}{Phys. Rev. Lett.} \textbf{\bibinfo{volume}{112}},
  \bibinfo{pages}{016401} (\bibinfo{year}{2014}).

\bibitem[{\citenamefont{De~Liberato et~al.}(2007)\citenamefont{De~Liberato,
  Ciuti, and Carusotto}}]{LiberatoetAl07PRL}
\bibinfo{author}{\bibfnamefont{S.}~\bibnamefont{De~Liberato}},
  \bibinfo{author}{\bibfnamefont{C.}~\bibnamefont{Ciuti}}, \bibnamefont{and}
  \bibinfo{author}{\bibfnamefont{I.}~\bibnamefont{Carusotto}},
  \bibinfo{journal}{Phys. Rev. Lett.} \textbf{\bibinfo{volume}{98}},
  \bibinfo{pages}{103602} (\bibinfo{year}{2007}).

\bibitem[{\citenamefont{Moore}(1970)}]{Moore70JMP}
\bibinfo{author}{\bibfnamefont{G.~T.} \bibnamefont{Moore}},
  \bibinfo{journal}{J. Math. Phys.} \textbf{\bibinfo{volume}{11}},
  \bibinfo{pages}{2679} (\bibinfo{year}{1970}).

\bibitem[{\citenamefont{Fulling and Davies}(1976)}]{FullingDavies76PRSLA}
\bibinfo{author}{\bibfnamefont{S.~A.} \bibnamefont{Fulling}} \bibnamefont{and}
  \bibinfo{author}{\bibfnamefont{P.~C.~W.} \bibnamefont{Davies}},
  \bibinfo{journal}{Proc. Roy. Soc. London A} \textbf{\bibinfo{volume}{348}},
  \bibinfo{pages}{393} (\bibinfo{year}{1976}).

\bibitem[{\citenamefont{Imamo\u{g}lu et~al.}(1999)\citenamefont{Imamo\u{g}lu,
  Awschalom, Burkard, DiVincenzo, Loss, Sherwin, and
  Small}}]{ImamogluetAl99PRL}
\bibinfo{author}{\bibfnamefont{A.}~\bibnamefont{Imamo\u{g}lu}},
  \bibinfo{author}{\bibfnamefont{D.~D.} \bibnamefont{Awschalom}},
  \bibinfo{author}{\bibfnamefont{G.}~\bibnamefont{Burkard}},
  \bibinfo{author}{\bibfnamefont{D.~P.} \bibnamefont{DiVincenzo}},
  \bibinfo{author}{\bibfnamefont{D.}~\bibnamefont{Loss}},
  \bibinfo{author}{\bibfnamefont{M.}~\bibnamefont{Sherwin}}, \bibnamefont{and}
  \bibinfo{author}{\bibfnamefont{A.}~\bibnamefont{Small}},
  \bibinfo{journal}{Phys. Rev. Lett.} \textbf{\bibinfo{volume}{83}},
  \bibinfo{pages}{4204} (\bibinfo{year}{1999}).

\bibitem[{\citenamefont{Blais et~al.}(2004)\citenamefont{Blais, Huang,
  Wallraff, Girvin, and Schoelkopf}}]{BlaisetAl04PRA}
\bibinfo{author}{\bibfnamefont{A.}~\bibnamefont{Blais}},
  \bibinfo{author}{\bibfnamefont{R.-S.} \bibnamefont{Huang}},
  \bibinfo{author}{\bibfnamefont{A.}~\bibnamefont{Wallraff}},
  \bibinfo{author}{\bibfnamefont{S.~M.} \bibnamefont{Girvin}},
  \bibnamefont{and} \bibinfo{author}{\bibfnamefont{R.~J.}
  \bibnamefont{Schoelkopf}}, \bibinfo{journal}{Phys. Rev. A}
  \textbf{\bibinfo{volume}{69}}, \bibinfo{pages}{062320}
  (\bibinfo{year}{2004}).

\bibitem[{\citenamefont{Todorov et~al.}(2009)\citenamefont{Todorov, Andrews,
  Sagnes, Colombelli, Klang, Strasser, and Sirtori}}]{TodorovetAl09PRL}
\bibinfo{author}{\bibfnamefont{Y.}~\bibnamefont{Todorov}},
  \bibinfo{author}{\bibfnamefont{A.}~\bibnamefont{Andrews}},
  \bibinfo{author}{\bibfnamefont{I.}~\bibnamefont{Sagnes}},
  \bibinfo{author}{\bibfnamefont{R.}~\bibnamefont{Colombelli}},
  \bibinfo{author}{\bibfnamefont{P.}~\bibnamefont{Klang}},
  \bibinfo{author}{\bibfnamefont{G.}~\bibnamefont{Strasser}}, \bibnamefont{and}
  \bibinfo{author}{\bibfnamefont{C.}~\bibnamefont{Sirtori}},
  \bibinfo{journal}{Phys. Rev. Lett.} \textbf{\bibinfo{volume}{102}},
  \bibinfo{pages}{186402} (\bibinfo{year}{2009}).

\bibitem[{\citenamefont{Geiser et~al.}(2012)\citenamefont{Geiser, Castellano,
  Scalari, Beck, Nevou, and Faist}}]{GeiseretAl12PRL}
\bibinfo{author}{\bibfnamefont{M.}~\bibnamefont{Geiser}},
  \bibinfo{author}{\bibfnamefont{F.}~\bibnamefont{Castellano}},
  \bibinfo{author}{\bibfnamefont{G.}~\bibnamefont{Scalari}},
  \bibinfo{author}{\bibfnamefont{M.}~\bibnamefont{Beck}},
  \bibinfo{author}{\bibfnamefont{L.}~\bibnamefont{Nevou}}, \bibnamefont{and}
  \bibinfo{author}{\bibfnamefont{J.}~\bibnamefont{Faist}},
  \bibinfo{journal}{Phys. Rev. Lett.} \textbf{\bibinfo{volume}{108}},
  \bibinfo{pages}{106402} (\bibinfo{year}{2012}).

\bibitem[{\citenamefont{Scalari et~al.}(2012)\citenamefont{Scalari, Maissen,
  Tur{\v{c}}inkov{\'a}, Hagenm\"uller, De~Liberato, Ciuti, Reichl, Schuh,
  Wegscheider, Beck et~al.}}]{ScalarietAl12Science}
\bibinfo{author}{\bibfnamefont{G.}~\bibnamefont{Scalari}},
  \bibinfo{author}{\bibfnamefont{C.}~\bibnamefont{Maissen}},
  \bibinfo{author}{\bibfnamefont{D.}~\bibnamefont{Tur{\v{c}}inkov{\'a}}},
  \bibinfo{author}{\bibfnamefont{D.}~\bibnamefont{Hagenm\"uller}},
  \bibinfo{author}{\bibfnamefont{S.}~\bibnamefont{De~Liberato}},
  \bibinfo{author}{\bibfnamefont{C.}~\bibnamefont{Ciuti}},
  \bibinfo{author}{\bibfnamefont{C.}~\bibnamefont{Reichl}},
  \bibinfo{author}{\bibfnamefont{D.}~\bibnamefont{Schuh}},
  \bibinfo{author}{\bibfnamefont{W.}~\bibnamefont{Wegscheider}},
  \bibinfo{author}{\bibfnamefont{M.}~\bibnamefont{Beck}}, \bibnamefont{et~al.},
  \bibinfo{journal}{Science} \textbf{\bibinfo{volume}{335}},
  \bibinfo{pages}{1323} (\bibinfo{year}{2012}).

\bibitem[{\citenamefont{Maissen et~al.}(2014)\citenamefont{Maissen, Scalari,
  Valmorra, Beck, Faist, Cibella, Leoni, Reichl, Charpentier, and
  Wegscheider}}]{MaissenetAl14PRB}
\bibinfo{author}{\bibfnamefont{C.}~\bibnamefont{Maissen}},
  \bibinfo{author}{\bibfnamefont{G.}~\bibnamefont{Scalari}},
  \bibinfo{author}{\bibfnamefont{F.}~\bibnamefont{Valmorra}},
  \bibinfo{author}{\bibfnamefont{M.}~\bibnamefont{Beck}},
  \bibinfo{author}{\bibfnamefont{J.}~\bibnamefont{Faist}},
  \bibinfo{author}{\bibfnamefont{S.}~\bibnamefont{Cibella}},
  \bibinfo{author}{\bibfnamefont{R.}~\bibnamefont{Leoni}},
  \bibinfo{author}{\bibfnamefont{C.}~\bibnamefont{Reichl}},
  \bibinfo{author}{\bibfnamefont{C.}~\bibnamefont{Charpentier}},
  \bibnamefont{and}
  \bibinfo{author}{\bibfnamefont{W.}~\bibnamefont{Wegscheider}},
  \bibinfo{journal}{Phys. Rev. B} \textbf{\bibinfo{volume}{90}},
  \bibinfo{pages}{205309} (\bibinfo{year}{2014}).

\bibitem[{\citenamefont{Haroche}(2013)}]{Haroche13RMP}
\bibinfo{author}{\bibfnamefont{S.}~\bibnamefont{Haroche}},
  \bibinfo{journal}{Rev. Mod. Phys.} \textbf{\bibinfo{volume}{85}},
  \bibinfo{pages}{1083} (\bibinfo{year}{2013}).

\bibitem[{\citenamefont{Khitrova et~al.}(2006)\citenamefont{Khitrova, Gibbs,
  Kira, Koch, and Scherer}}]{KhitrovaetAl06NP}
\bibinfo{author}{\bibfnamefont{G.}~\bibnamefont{Khitrova}},
  \bibinfo{author}{\bibfnamefont{H.~M.} \bibnamefont{Gibbs}},
  \bibinfo{author}{\bibfnamefont{M.}~\bibnamefont{Kira}},
  \bibinfo{author}{\bibfnamefont{S.~W.} \bibnamefont{Koch}}, \bibnamefont{and}
  \bibinfo{author}{\bibfnamefont{A.}~\bibnamefont{Scherer}},
  \bibinfo{journal}{Nat. Phys.} \textbf{\bibinfo{volume}{2}},
  \bibinfo{pages}{81} (\bibinfo{year}{2006}).

\bibitem[{\citenamefont{T\"orm\"a and Barnes}(2015)}]{TormaBarnes15RPP}
\bibinfo{author}{\bibfnamefont{P.}~\bibnamefont{T\"orm\"a}} \bibnamefont{and}
  \bibinfo{author}{\bibfnamefont{W.~L.} \bibnamefont{Barnes}},
  \bibinfo{journal}{Rep. Prog. Phys.} \textbf{\bibinfo{volume}{78}},
  \bibinfo{pages}{013901} (\bibinfo{year}{2015}).

\bibitem[{\citenamefont{Tabuchi et~al.}(2015)\citenamefont{Tabuchi, Ishino,
  Noguchi, Ishikawa, Yamazaki, Usami, and Nakamura}}]{TabuchietAl15Science}
\bibinfo{author}{\bibfnamefont{Y.}~\bibnamefont{Tabuchi}},
  \bibinfo{author}{\bibfnamefont{S.}~\bibnamefont{Ishino}},
  \bibinfo{author}{\bibfnamefont{A.}~\bibnamefont{Noguchi}},
  \bibinfo{author}{\bibfnamefont{T.}~\bibnamefont{Ishikawa}},
  \bibinfo{author}{\bibfnamefont{R.}~\bibnamefont{Yamazaki}},
  \bibinfo{author}{\bibfnamefont{K.}~\bibnamefont{Usami}}, \bibnamefont{and}
  \bibinfo{author}{\bibfnamefont{Y.}~\bibnamefont{Nakamura}},
  \bibinfo{journal}{Science} \textbf{\bibinfo{volume}{349}},
  \bibinfo{pages}{405} (\bibinfo{year}{2015}).

\bibitem[{\citenamefont{Viennot et~al.}(2015)\citenamefont{Viennot, Dartiailh,
  Cottet, and Kontos}}]{ViennotetAl15Science}
\bibinfo{author}{\bibfnamefont{J.~J.} \bibnamefont{Viennot}},
  \bibinfo{author}{\bibfnamefont{M.~C.} \bibnamefont{Dartiailh}},
  \bibinfo{author}{\bibfnamefont{A.}~\bibnamefont{Cottet}}, \bibnamefont{and}
  \bibinfo{author}{\bibfnamefont{T.}~\bibnamefont{Kontos}},
  \bibinfo{journal}{Science} \textbf{\bibinfo{volume}{349}},
  \bibinfo{pages}{408} (\bibinfo{year}{2015}).

\bibitem[{\citenamefont{Dicke}(1954)}]{Dicke54PR}
\bibinfo{author}{\bibfnamefont{R.~H.} \bibnamefont{Dicke}},
  \bibinfo{journal}{Phys. Rev.} \textbf{\bibinfo{volume}{93}},
  \bibinfo{pages}{99} (\bibinfo{year}{1954}).

\bibitem[{\citenamefont{Bonifacio and Lugiato}(1982)}]{BonifacioLugiato82Book}
\bibinfo{author}{\bibfnamefont{R.}~\bibnamefont{Bonifacio}} \bibnamefont{and}
  \bibinfo{author}{\bibfnamefont{L.~A.} \bibnamefont{Lugiato}}, in
  \emph{\bibinfo{booktitle}{Dissipative Systems in Quantum Optics}}, edited by
  \bibinfo{editor}{\bibfnamefont{R.}~\bibnamefont{Bonifacio}}
  (\bibinfo{publisher}{Springer-Verlag}, \bibinfo{address}{Berlin},
  \bibinfo{year}{1982}), Topics in Current Physics, chap.~\bibinfo{chapter}{4},
  pp. \bibinfo{pages}{61--92}.

\bibitem[{\citenamefont{Deng et~al.}(2010)\citenamefont{Deng, Haug, and
  Yamamoto}}]{DengetAl10RMP}
\bibinfo{author}{\bibfnamefont{H.}~\bibnamefont{Deng}},
  \bibinfo{author}{\bibfnamefont{H.}~\bibnamefont{Haug}}, \bibnamefont{and}
  \bibinfo{author}{\bibfnamefont{Y.}~\bibnamefont{Yamamoto}},
  \bibinfo{journal}{Rev. Mod. Phys.} \textbf{\bibinfo{volume}{82}},
  \bibinfo{pages}{1489} (\bibinfo{year}{2010}).

\bibitem[{\citenamefont{Hagenm{\"u}ller
  et~al.}(2010)\citenamefont{Hagenm{\"u}ller, De~Liberato, and
  Ciuti}}]{HagenmulleretAl10PRB}
\bibinfo{author}{\bibfnamefont{D.}~\bibnamefont{Hagenm{\"u}ller}},
  \bibinfo{author}{\bibfnamefont{S.}~\bibnamefont{De~Liberato}},
  \bibnamefont{and} \bibinfo{author}{\bibfnamefont{C.}~\bibnamefont{Ciuti}},
  \bibinfo{journal}{Phys. Rev. B} \textbf{\bibinfo{volume}{81}},
  \bibinfo{pages}{235303} (\bibinfo{year}{2010}).

\bibitem[{\citenamefont{Kaluzny et~al.}(1983)\citenamefont{Kaluzny, Goy, Gross,
  Raimond, and Haroche}}]{KaluznyetAl83PRL}
\bibinfo{author}{\bibfnamefont{Y.}~\bibnamefont{Kaluzny}},
  \bibinfo{author}{\bibfnamefont{P.}~\bibnamefont{Goy}},
  \bibinfo{author}{\bibfnamefont{M.}~\bibnamefont{Gross}},
  \bibinfo{author}{\bibfnamefont{J.~M.} \bibnamefont{Raimond}},
  \bibnamefont{and} \bibinfo{author}{\bibfnamefont{S.}~\bibnamefont{Haroche}},
  \bibinfo{journal}{Phys. Rev. Lett.} \textbf{\bibinfo{volume}{51}},
  \bibinfo{pages}{1175} (\bibinfo{year}{1983}).

\bibitem[{\citenamefont{Ams\"uss et~al.}(2011)\citenamefont{Ams\"uss, Koller,
  N\"obauer, Putz, Rotter, Sandner, Schneider, Schramb\"ock, Steinhauser,
  Ritsch et~al.}}]{AmsussetAl11PRL}
\bibinfo{author}{\bibfnamefont{R.}~\bibnamefont{Ams\"uss}},
  \bibinfo{author}{\bibfnamefont{C.}~\bibnamefont{Koller}},
  \bibinfo{author}{\bibfnamefont{T.}~\bibnamefont{N\"obauer}},
  \bibinfo{author}{\bibfnamefont{S.}~\bibnamefont{Putz}},
  \bibinfo{author}{\bibfnamefont{S.}~\bibnamefont{Rotter}},
  \bibinfo{author}{\bibfnamefont{K.}~\bibnamefont{Sandner}},
  \bibinfo{author}{\bibfnamefont{S.}~\bibnamefont{Schneider}},
  \bibinfo{author}{\bibfnamefont{M.}~\bibnamefont{Schramb\"ock}},
  \bibinfo{author}{\bibfnamefont{G.}~\bibnamefont{Steinhauser}},
  \bibinfo{author}{\bibfnamefont{H.}~\bibnamefont{Ritsch}},
  \bibnamefont{et~al.}, \bibinfo{journal}{Phys. Rev. Lett.}
  \textbf{\bibinfo{volume}{107}}, \bibinfo{pages}{060502}
  (\bibinfo{year}{2011}).

\bibitem[{\citenamefont{Tabuchi et~al.}(2014)\citenamefont{Tabuchi, Ishino,
  Ishikawa, Yamazaki, Usami, and Nakamura}}]{TabuchietAl14PRL}
\bibinfo{author}{\bibfnamefont{Y.}~\bibnamefont{Tabuchi}},
  \bibinfo{author}{\bibfnamefont{S.}~\bibnamefont{Ishino}},
  \bibinfo{author}{\bibfnamefont{T.}~\bibnamefont{Ishikawa}},
  \bibinfo{author}{\bibfnamefont{R.}~\bibnamefont{Yamazaki}},
  \bibinfo{author}{\bibfnamefont{K.}~\bibnamefont{Usami}}, \bibnamefont{and}
  \bibinfo{author}{\bibfnamefont{Y.}~\bibnamefont{Nakamura}},
  \bibinfo{journal}{Phys. Rev. Lett.} \textbf{\bibinfo{volume}{113}},
  \bibinfo{pages}{083603} (\bibinfo{year}{2014}).

\bibitem[{\citenamefont{Zhang et~al.}(2014{\natexlab{a}})\citenamefont{Zhang,
  Zou, Jiang, and Tang}}]{ZhangetAl14PRL2}
\bibinfo{author}{\bibfnamefont{X.}~\bibnamefont{Zhang}},
  \bibinfo{author}{\bibfnamefont{C.-L.} \bibnamefont{Zou}},
  \bibinfo{author}{\bibfnamefont{L.}~\bibnamefont{Jiang}}, \bibnamefont{and}
  \bibinfo{author}{\bibfnamefont{H.~X.} \bibnamefont{Tang}},
  \bibinfo{journal}{Phys. Rev. Lett.} \textbf{\bibinfo{volume}{113}},
  \bibinfo{pages}{156401} (\bibinfo{year}{2014}{\natexlab{a}}).

\bibitem[{\citenamefont{Zhang et~al.}(2014{\natexlab{b}})\citenamefont{Zhang,
  Arikawa, Kato, Reno, Pan, Watson, Manfra, Zudov, Tokman, Erukhimova
  et~al.}}]{ZhangetAl14PRL}
\bibinfo{author}{\bibfnamefont{Q.}~\bibnamefont{Zhang}},
  \bibinfo{author}{\bibfnamefont{T.}~\bibnamefont{Arikawa}},
  \bibinfo{author}{\bibfnamefont{E.}~\bibnamefont{Kato}},
  \bibinfo{author}{\bibfnamefont{J.~L.} \bibnamefont{Reno}},
  \bibinfo{author}{\bibfnamefont{W.}~\bibnamefont{Pan}},
  \bibinfo{author}{\bibfnamefont{J.~D.} \bibnamefont{Watson}},
  \bibinfo{author}{\bibfnamefont{M.~J.} \bibnamefont{Manfra}},
  \bibinfo{author}{\bibfnamefont{M.~A.} \bibnamefont{Zudov}},
  \bibinfo{author}{\bibfnamefont{M.}~\bibnamefont{Tokman}},
  \bibinfo{author}{\bibfnamefont{M.}~\bibnamefont{Erukhimova}},
  \bibnamefont{et~al.}, \bibinfo{journal}{Phys. Rev. Lett.}
  \textbf{\bibinfo{volume}{113}}, \bibinfo{pages}{047601}
  (\bibinfo{year}{2014}{\natexlab{b}}).

\bibitem[{\citenamefont{Norris et~al.}(1994)\citenamefont{Norris, Rhee, Sung,
  Arakawa, Nishioka, and Weisbuch}}]{NorrisetAl94PRB}
\bibinfo{author}{\bibfnamefont{T.~B.} \bibnamefont{Norris}},
  \bibinfo{author}{\bibfnamefont{J.-K.} \bibnamefont{Rhee}},
  \bibinfo{author}{\bibfnamefont{C.-Y.} \bibnamefont{Sung}},
  \bibinfo{author}{\bibfnamefont{Y.}~\bibnamefont{Arakawa}},
  \bibinfo{author}{\bibfnamefont{M.}~\bibnamefont{Nishioka}}, \bibnamefont{and}
  \bibinfo{author}{\bibfnamefont{C.}~\bibnamefont{Weisbuch}},
  \bibinfo{journal}{Phys. Rev. B} \textbf{\bibinfo{volume}{50}},
  \bibinfo{pages}{14663} (\bibinfo{year}{1994}).

\bibitem[{\citenamefont{Tignon et~al.}(1995)\citenamefont{Tignon, Voisin,
  Delalande, Voos, Houdr\'e, Oesterle, and Stanley}}]{TignonetAl95PRL}
\bibinfo{author}{\bibfnamefont{J.}~\bibnamefont{Tignon}},
  \bibinfo{author}{\bibfnamefont{P.}~\bibnamefont{Voisin}},
  \bibinfo{author}{\bibfnamefont{C.}~\bibnamefont{Delalande}},
  \bibinfo{author}{\bibfnamefont{M.}~\bibnamefont{Voos}},
  \bibinfo{author}{\bibfnamefont{R.}~\bibnamefont{Houdr\'e}},
  \bibinfo{author}{\bibfnamefont{U.}~\bibnamefont{Oesterle}}, \bibnamefont{and}
  \bibinfo{author}{\bibfnamefont{R.~P.} \bibnamefont{Stanley}},
  \bibinfo{journal}{Phys. Rev. Lett.} \textbf{\bibinfo{volume}{74}},
  \bibinfo{pages}{3967} (\bibinfo{year}{1995}).

\bibitem[{\citenamefont{Andreev et~al.}(2014)\citenamefont{Andreev, Muravev,
  Belyanin, and Kukushkin}}]{AndreevetAl14APL}
\bibinfo{author}{\bibfnamefont{I.~V.} \bibnamefont{Andreev}},
  \bibinfo{author}{\bibfnamefont{V.~M.} \bibnamefont{Muravev}},
  \bibinfo{author}{\bibfnamefont{V.~N.} \bibnamefont{Belyanin}},
  \bibnamefont{and} \bibinfo{author}{\bibfnamefont{I.~V.}
  \bibnamefont{Kukushkin}}, \bibinfo{journal}{Appl. Phys. Lett.}
  \textbf{\bibinfo{volume}{105}}, \bibinfo{pages}{202106}
  (\bibinfo{year}{2014}).

\end{thebibliography}


\end{document}